# Interaction Driven Quantum Hall Wedding cake-like Structures in Graphene Quantum Dots

Christopher Gutiérrez[1,2,*,†], Daniel Walkup[1,2,*], Fereshte Ghahari[1,2,*], Cyprian Lewandowski[3,*], Joaquin F. Rodriguez-Nieva[4], Kenji Watanabe[5], Takashi Taniguchi[5], Leonid S. Levitov[3], Nikolai B. Zhitenev[1], and Joseph A. Stroscio[1,‡]

[1]Center for Nanoscale Science and Technology, National Institute of Standards and Technology, Gaithersburg, MD 20899, USA
[2]Maryland NanoCenter, University of Maryland, College Park, MD 20742, USA
[3]Department of Physics, Massachusetts Institute of Technology, Cambridge, MA 02139, USA
[4]Department of Physics, Harvard University, Cambridge, MA 02138, USA
[5]National Institute for Materials Science, Tsukuba, Ibaraki 305-0044, Japan

**Quantum-relativistic matter is ubiquitous in nature; however it is notoriously difficult to probe. The ease with which external electric and magnetic fields can be introduced in graphene opens a door to creating a table-top prototype of strongly confined relativistic matter. Here, through a detailed spectroscopic mapping, we provide a spatial visualization of the interplay between spatial and magnetic confinement in a circular graphene resonator. We directly observe the development of a multi-tiered "wedding cake"-like structure of concentric regions of compressible/incompressible quantum Hall states, a signature of electron interactions in the system. Solid-state experiments can therefore yield insights into the behaviour of quantum-relativistic matter under extreme conditions.**

---

[*] These authors contributed equally to this work.
[†] Present address: Quantum Matter Institute, University of British Columbia, Vancouver, British Columbia V6T 1Z4, Canada
[‡] To whom correspondence should be addressed.

Energy quantization due to quantum confinement takes place when the particle's de Broglie wavelength becomes comparable to the system's length scale. Confinement can arise through spatial constraints imposed by electric fields or through cyclotron motion induced by magnetic fields. Combined together, confinement and quantization strengthen the effects of electron-electron interactions, providing a setting to probe a range of exotic phenomena in strongly correlated quantum systems. In the solid-state setting different types of confined strongly correlated states and transitions between them have been studied in quantum dots (QD) in the presence of external magnetic fields (*1*). Evolution from atomic-like shell structure to magnetic quantization in QDs was first probed using Coulomb blockade spectroscopy (*2–4*). For QDs at large magnetic fields, *i.e.* in the quantum Hall regime, it is expected that Coulomb interactions and the redistribution of carriers between Landau levels (LLs) will lead to a characteristic wedding cake-like shape in the density of electronic states (Fig. 1E) (*4–8*). Although similar structures have been observed in ultracold atoms undergoing transition from the superfluid to Mott insulator (*9, 10*), they have not been mapped spatially in a solid-state system.

Graphene offers an ideal platform for this enquiry as it hosts a fully exposed two-dimensional electron gas amenable to local probes (*11–17*). Graphene circular *p-n* junction resonators (*18–22*) with built-in local potentials (Fig. 1A) are particularly well suited to the present study; they circumvent the problems of edge roughness and edge impurities encountered in lithographically fabricated graphene QDs. Further, they enable fine control of the confining potential as well as QD doping by means of local gate potentials, offering an opportunity to directly visualize the transition of electron states from spatial to magnetic confinement (Fig. 1, B to E).

In the absence of a magnetic field, confinement of graphene carriers in a *p-n* junction resonator gives rise to a series of quasi-bound single particle states. These states result from oblique Klein scattering at the *p-n* interface (*18–22*). At the same time, Klein tunneling, although present, is very weak for oblique scattering angles and thus has little impact on confinement (*23*). Analogous to atomic physics, the many-electron shell-like states are characterized by radial and azimuthal quantum numbers $(n, m)$, forming a ladder of states within the spatially confined potential (Fig. 1B) (*20, 22*). In weak magnetic fields, there is a giant splitting of energy levels corresponding to time-reversed $\pm m$ states induced by the $\pi$ Berry phase in graphene (*22, 24*). At higher fields, the system enters the quantum Hall regime, with confined states transitioning to highly degenerate LLs (Fig. 1, C and D). A signature of the transition is formation in the electron density of wedding cake-like structures comprised of a series of compressible and incompressible electron liquid rings (Fig. 1E) (*4–8*). Extending the single-particle description to include Coulomb interactions is essential in this regime.

Our experiment involves spectroscopic mapping of a graphene QD by tunneling measurements. The QD is formed by ionized impurities in the hexagonal boron nitride (hBN) insulating layer acting as a fixed built-in confining potential (see Fig. 1A and Refs. (*22*) and (*25*) for device fabrication). The transition from spatial to magnetic confinement occurs when the magnetic length becomes smaller than the confining potential width. By following the bright envelope in the spectral map in Fig. 2A at zero field, we can estimate the effective confining potential as $V_{B=0}(r) \approx U_0 \exp\left(-\frac{r^2}{R_0^2}\right) + U_\infty$, with $U_0 \approx 210$ meV, $R_0 \approx 104$ nm and $U_\infty = -55$ meV. This defines a characteristic length scale for the confining potential $l_V = \left(\frac{R_0^2 \hbar v_F}{U_0}\right)^{1/3} \approx 32$ nm. Here $\hbar$ is Plank's constant, $e$ is the elementary charge, and $v_F \approx 10^6$ m/s is the graphene Fermi velocity. Such a potential gives rise to quasi-bound states with energy splitting $\Delta E \approx (\hbar v_F)/l_V \approx 20$ meV.

Application of a magnetic field tends to confine electrons in a region of size $l_B = \left(\frac{\hbar}{eB}\right)^{1/2}$ and leads to the characteristic Landau quantization in graphene, $\varepsilon_N = \text{sgn}(N)\hbar v_F \sqrt{2|N|}/l_B$, with $N = \pm 0, \pm 1, \pm 2, \ldots$. Each LL is highly degenerate and can host $n_{LL} = g/2\pi l_B^2$ carriers per unit area, where $g = 4$ is the valley/spin degeneracy. Therefore, we expect a transition from atomic-like QD states to LL states occurring at $l_B \cong l_V$, which gives a transitional magnetic field of $B \approx 0.6$ T.

The transition and intricate evolution of QD states from spatial to magnetic confinement with increasing magnetic field is displayed in Fig. 2. The measured differential conductance signal, proportional to the local density of states (LDOS), shows the evolution of the QD states in the energy vs radial plane that cuts through the diameter of the QD. The zero-field shell-like QD states are well-resolved in Fig. 2A under the bright concave band, which follows the confining potential. States with $(m = \pm\frac{1}{2})$ have the largest weight in the center at $r = 0$, whereas states with a common radial quantum number $n$, have a large weight in the form of arcs following the concave potential outline. The first critical field is reached by 0.25 T, where the $\pm m$ degeneracy is lifted owing to the turn on of a $\pi$ Berry phase (*22*, *24*), as seen by the doubling of the anti-nodes at $r = 0$ (arrows in Fig. 2B). The onset of the transition into the quantum Hall regime can be observed at $B = 0.5$ T (Fig. 2C) in agreement with the estimate above. Here states in the center of the resonator start to flatten out, have increased intensity, and shift lower in energy. Beginning at $B = 1$ T various interior resonator states (arrows in Fig. 2D) merge into the $N = 0$ Landau level (LL(0)). With progressively higher fields, the number of QD resonances decreases as they condense into the flattened central states forming a series of highly degenerate LLs (Fig. 2, F to I). Beginning at about 2 T (Fig. 2F), LL(0) develops kinks near the QD boundary and an additional concave cusp near the center. Below we argue that these are related to electron interaction effects. Additionally, LL(0) develops a splitting, which increases with field, whereas LL(-1) continually moves down in energy. In this field range, a decrease in conductance over a small energy range is also observed at the Fermi level, which we attribute to a Coulomb pseudogap (*26*, *27*).

We now discuss the spatial pattern associated with the eigenstate evolution observed in Fig. 2. Experimental spatial maps of the differential conductance, corresponding to resonator LDOS wavefunction probability amplitudes, are obtained by taking a two-dimensional slice in the *x-y* plane of the data set in Fig. 2 at a specific energy (Fig. 3). Only a subset of the data is shown in Fig. 3 corresponding to specific energies of the prominent central states at $r = 0$ in Fig. 2, with increasing magnetic field for each column of maps [a complete view of the data set can be seen in the movie file S1]. The spatial extent of the $m=\pm1/2$ states at the selected energies is observed at zero field in the first column of the maps in Fig. 3. As we increase field and progress from the left to the right of the figure, we observe the formation of rings with diameters that narrow both in diameter and width with increasing field. For higher *n* states (progressing down in a column), more rings are seen. Some of these rings originate from the quasi-bound resonances that have not yet developed into Landau levels and some reflect the presence of magnetic confinement. The former can have a relatively narrow spatial profile if they are dispersing up or down in energy. At a still higher field (Fig. 3D) LLs plateaus are formed as seen in Fig. 2 and show up as bright rings in the spatial maps as indicated by the arrows pointing to LL(+1) and the valley-split (*28*) LL(0+), LL(0-) state in Fig. 3D. When LL states cross, or are pinned at the Fermi level, they form compressible (metallic) rings and disks, which start to show Coulomb charging effects (*13*), as indicated by the fine quartet of rings in the center and outside edge of Fig. 3D (see also vertically dispersing lines in Fig. 4E and rings in Fig 4G).

A striking and unexpected feature observed in Fig. 3 is the appearance of circular nodal patterns in the spatial maps of differential conductance, which are present even at zero field. The origin of these nodal patterns is not clear at present, but they can either be attributed to interactions as they resemble the shell-like structure predicted for Wigner crystals (*7, 29–31*) or to deviations from a rotationally symmetric confining potential. Deviations from perfect symmetry will partially lift the *m*-state degeneracy and give rise to nodal patterns. Moiré superlattice effects can be ruled out as an origin of the potential asymmetry because of the large angular mismatch (≈29 °) between the graphene and hBN insulator for this device, which gives a superlattice period of ≈0.5 nm, much smaller than the nodal separation length scales (*22*). A non-symmetric potential can result from the shape of the probe tip, which gets imprinted in the QD potential shape from the electric field generated during the tip voltage pulse.

As a simple theoretical model, we use the edge state picture of the quantum Hall effect. In a circular geometry, it yields a system of compressible and incompressible rings formed in the electron liquid (Fig. 1D) owing to the interplay between Landau levels and electron interactions (*4*). In our measurement, interaction effects are observed already at low fields, signaled by the shifting and flattening of the LL states in Fig. 2. A minimal model incorporating interactions at low fields is the energy functional (*4*)

$$E[n] = \int d^2r \left( K[n(r)] + V_{\text{ext}}(r)n(r) + \frac{1}{2}\int d^2r' V_{ee}(|\mathbf{r} - \mathbf{r}'|)n(r)n(r') \right), \quad (1)$$

where $n(r)$ is charge density at position $\mathbf{r}$. This functional describes the competition between carriers' kinetic energy $K[n(r)]$ and the effective potential $V_B(r)$. We approximate these quantities as:

$$\frac{\delta K}{\delta n(r)} = \varepsilon_N, \quad \left(N - \frac{1}{2}\right)n_{\text{LL}} < n(r) < \left(N + \frac{1}{2}\right)n_{\text{LL}}, \quad (2)$$

$$V_B(r) = V_{\text{ext}}(r) + \int d^2r' V_{ee}(|\mathbf{r} - \mathbf{r}'|)n(r'). \quad (3)$$

These relations are valid in the limit $l_B \ll l_V$. Here $N = 0, \pm 1, \pm 2 \ldots$ is the LL number, $V_{\text{ext}}$ is the electrostatic potential defining the dot, $V_{ee}(r) = \frac{\tilde{e}^2}{r}$ is the Coulomb interaction and $\tilde{e}$ is the screened electron charge (see (*25*) for details).

The calculated effective potential $V_B(r)$ is shown in Fig. 4A for a few magnetic field values. The joint effect of the magnetic field and interactions is to create a series of plateaus forming a multi-tiered wedding cake-like pattern of concentric rings within the dot. At the same time the potential is reduced compared to $V_{B=0}(r)$ due to screening. Notably, the reduced potential causes LL(0) to move toward the Fermi level in agreement with the energy dependence of LL(0) in Fig. 2. The extra concave features in the potential in the central region match those in the experimental maps of LL(0) in Fig. 2, F to I. The effect of interactions on the LLs is shown by comparing the LDOS with and without interactions in the left and right panels of Fig. 4F. Before interactions are turned on (right panel), the LLs seen through the LDOS essentially track the potential $V_{B=0}(r)$. After including interactions (left panel), the evolution of the LDOS mimics that of potential $V_B(r)$: LLs shift to lower energy and flatten in the central region, in agreement with the evolution seen in the measurements (Fig. 2).

The incompressible and compressible rings become considerably clearer in higher magnetic field. The experimental spectral map in Fig. 4E shows the LLs becoming flat in the central region of the QD even though the bare external electrostatic potential is concave (see QD outline Fig. 2A), and then they progress sharply to a new energy level as new LLs become occupied, forming a wedding cake-like structure. Here LL($N$), $N = -5$ to 2 can be observed as plateaus in the center of the QD (Fig. 4E). Both LL(0) and LL(-1) cross the Fermi level at zero bias as indicated by the yellow lines, forming a LL(-1) compressible disk in the center and an outer LL(0) compressible ring separated by an incompressible ring, as shown in the Fermi-level spatial map in the *x-y* plane (Fig. 4G). We observe Coulomb charging of these LLs as charging lines intersecting the LLs at the Fermi level and progressing upward at sharp angles in Fig. 4E. These lines correspond to a quartet of rings in Fig. 4G and Fig. 3D. The charging of the compressible regions occurs in groups of four, reflecting the four-fold (spin and valley) graphene degeneracy (*13*).

To understand these observations, we use a two-stage approach. We first use the mean field functional Eq. 1 to find LL occupancies and determine the screened potential $V_B(r)$ (*25*). We then use this potential to calculate the density of microscopic states, which can be directly compared to the measurements. The features seen in the measured LDOS can be understood by comparing them to a simple calculation of the LLs (*25*), shown in Fig. 4C. The highest LL that is partially filled can be obtained by counting the number of LLs that need to be populated to accommodate the carrier density equal to that in the fully compressible regime (dashed line in Fig. 4B). In the simulated LDOS map in Fig. 4D, we can identify the LL states, which track the screened potential $V_B(r)$ pictured in Fig. 4A, and exhibit plateaus as expected from theory (*4*, *5*). This behavior is in good agreement with the experimental results shown in Fig. 4E.

The width of the observed incompressible ring can be estimated from the functional in Eq. 1 following the approach of Ref. (*5*) and yields the strip width (*25*),

$$l = \left( \frac{4\Delta\varepsilon_{\text{LL}}}{\pi^2 \tilde{e}^2 \frac{dn}{dr}} \right)^{\frac{1}{2}} \approx 34 \text{nm}. \quad (4)$$

The estimate in Eq. 4 is slightly greater than the width inferred from our measurement results shown in Fig. 4E. The small discrepancy can partly be attributed to the result of Ref. (*5*), derived for LL spacing $\Delta\varepsilon_{\text{LL}}$ much smaller than the external potential, being used in the regime when $\Delta\varepsilon_{\text{LL}}$ is not small on the $V_B(r)$ scale.

Fingerprints of electron-electron interactions that are as clear and striking as the observed electronic wedding cake-like patterns are relatively rare in solid-state experiments. The measurements reported here suggest, as hinted by the charging lines and nodal patterns in the differential conductance maps, that even more exotic signatures of electronic interactions may be within experimental grasp in future scanned probe measurements at lower temperatures.

**FIGURE CAPTIONS**

**Fig. 1 Schematic evolution of states with magnetic field in a graphene quantum dot.** (**A**) The device geometry for the graphene quantum dot resonator with a *p*-doped center inside a *n*-doped background used in the current experiment. (**B-E**), (top) Schematic of the potential profile (grey surface), and corresponding wavefunction density (orange surfaces), and (bottom) semi-classical orbits, as a function of applied magnetic field. Confined states start out as quasi-bound QD states and condense into LLs with increasing field. The corresponding screened potential develops a wedding cake-like appearance through electron interactions. Semi-classical orbits start out as expected for a central force potential and then develop into cyclotron motion drifting along equipotential lines forming compressible (blue) and incompressible (yellow) density rings, as shown in (E).

**Fig. 2 Visualization of the condensation of states from spatial to magnetic quantization.** (**A-I**) Experimental differential conductance ($T = 4.3$ K), $g(V_b, V_g, r, B)$, maps the local density of states as a function of applied magnetic field, showing manifolds of spatially confined QD states condensing into LLs at higher fields. The magnetic field and corresponding magnetic length $l_B$ is indicated in the bottom of each map. The 2D maps are radially averaged from a 2D grid of spectra. A smooth background was subtracted to remove the graphene dispersive background (*25*). The yellow arrows in (B) indicate the splitting of the $m = \pm 1/2$ degeneracy at $r = 0$ due to the turn on of a $\pi$ Berry phase. The blue arrows in (D) indicate the shell-like states merging into the *N*=0 Landau level edge mode. (**J**) Energy positions of the $(n, m = 1/2)$ states (symbols) obtained from the maps in (A-I) at $r = 0$ are observed to evolve into separate LLs with increasing applied magnetic field. LL(0) splits into two peaks above 2.5 T, indicated by the open and solid square symbols. The experimental uncertainty, determined from fitting the peak positions in the spectra, represents one standard deviation and is smaller than the symbol size.

**Fig. 3 Differential conductance spatial maps of QD states vs magnetic field**. Each column in the figure corresponds to differential conductance maps ($T$ = 4.3 K) in the *x-y* plane, $g(V_b, V_g, r, B)$, at a specific energy and at fixed magnetic fields from 0 T to 4 T, (**A-D**), respectively, indicated in the top of each column along with the magnetic length bar. The maps in each column are at energies corresponding to prominent QD states observed in Fig. 2, at the sample bias voltages indicated on top. With increased magnetic field, various circular rings appear to get narrower, reflecting the drift states schematically indicated in Fig. 1, B to E. Charging of compressible rings develops at the larger field in (D) evidenced by the quartets of fine rings in the center and outside edge (see Fig. 4E for the corresponding radial map). A smooth background was subtracted from each *dI/dV* vs $V_b$ spectra in the 2D grid to remove the graphene dispersive background (*25*).

**Fig. 4 Electron Interactions and the Wedding Cake-like Structure**. (**A**) Effective potential for several magnetic field values (solid lines) and the $V_{B=0}(r)$ fit based on Fig. 2A (dashed line). (**B-C**) Carrier density and Landau levels at *B*=4 T as predicted by the model from Eq. 1. Screening produces compressible regions, where LLs

are flat and pinned at the Fermi level, separated by incompressible regions (marked in grey). The size of incompressible regions is estimated in Eq. 4. The dashed line in (B) describes charge density in the compressible limit obtained by excluding the kinetic term $K[n(r)]$ from the functional in Eq. 1. (**D**) LDOS map calculated using the screened potential from (A). A Fermi velocity of $1.2 \times 10^6$ m/s was used in the calculations to match the LL positions between theory and experiment at $B = 4$ T. (**E**) Experimental differential conductance map ($T = 4.3$ K), $g(V_b, V_g, r, B)$, as a function of $V_b$ and $r$ at $B = 4$ T showing the wedding cake structure in the LLs in the QD. (**F**) LDOS simulated using the potential at $B=3$ T and $V_{B=0}(r)$ from (A) (see text for discussion). (**G**) An *x-y* slice of the $g(V_b, V_g, r, B)$ map in (F) at $V_b$=6 mV (near the Fermi level) showing the inner compressible disk from LL(-1) and the outer compressible ring from LL(0), as schematically indicated in Fig. 1E. The solid yellow lines show where LL(0) and LL(-1) cut through the Fermi level, creating the compressible rings.

**REFERENCES AND NOTES**

1. S. M. Reimann, M. Manninen, Electronic structure of quantum dots. *Rev. Mod. Phys.* **74**, 1283–1342 (2002).

2. R. C. Ashoori, Electrons in artificial atoms. *Nature*. **379**, 413–419 (1996).

3. L. Kouwenhoven *et al.*, Electron transport in quantum dots. *MESOSCOPIC ELECTRON Transp.* **345**, 105–214 (1997).

4. P. L. McEuen *et al.*, Self-consistent addition spectrum of a Coulomb island in the quantum Hall regime. *Phys. Rev. B*. **45**, 11419–11422 (1992).

5. D. Chklovskii, B. Shklovskii, L. Glazman, Electrostatics of edge channels. *Phys. Rev. B*. **46**, 4026–4034 (1992).

6. D. Chklovskii, K. Matveev, B. Shklovskii, Ballistic conductance of interacting electrons in the quantum Hall regime. *Phys. Rev. B*. **47**, 12605–12617 (1993).

7. Y. Nazarov, A. Khaetskii, Wigner molecule on the top of a quantum dot. *Phys. Rev. B*. **49**, 5077–5080 (1994).

8. M. Fogler, E. Levin, B. Shklovskii, Chemical potential and magnetization of a Coulomb island. *Phys. Rev. B*. **49**, 13767–13775 (1994).

9. S. Fölling, A. Widera, T. Müller, F. Gerbier, I. Bloch, Formation of Spatial Shell Structure in the Superfluid to Mott Insulator Transition. *Phys. Rev. Lett.* **97**, 060403 (2006).

10. N. Gemelke, X. Zhang, C.-L. Hung, C. Chin, In situ observation of incompressible Mott-insulating domains in ultracold atomic gases. *Nature*. **460**, 995–998 (2009).

11. J. Martin *et al.*, The nature of localization in graphene under quantum Hall conditions. *Nat. Phys.* **5**, 669–674 (2009).

12. D. L. Miller *et al.*, Observing the quantization of zero mass carriers in graphene. *Science*. **324**, 924–927 (2009).

13. S. Jung *et al.*, Evolution of microscopic localization in graphene in a magnetic field from scattering resonances to quantum dots. *Nat Phys*. **7**, 245–251 (2011).

14. J. Chae *et al.*, Renormalization of the Graphene Dispersion Velocity Determined from Scanning Tunneling Spectroscopy. *Phys. Rev. Lett.* **109**, 116802 (2012).

15. M. Zarenia, A. Chaves, G. A. Farias, F. M. Peeters, Energy levels of triangular and hexagonal graphene quantum dots: A comparative study between the tight-binding and Dirac equation approach. *Phys. Rev. B*. **84** (2011), doi:10.1103/PhysRevB.84.245403.

16. J.-B. Qiao *et al.*, Bound states in nanoscale graphene quantum dots in a continuous graphene sheet. *Phys. Rev. B*. **95**, 081409 (2017).

17. Y. Jiang *et al.*, Tuning a circular p–n junction in graphene from quantum confinement to optical guiding. *Nat. Nanotechnol.* **12**, 1045–1049 (2017).

18. Y. Zhao *et al.*, Creating and probing electron whispering-gallery modes in graphene. *Science*. **348**, 672–675 (2015).

19. N. M. Freitag *et al.*, Electrostatically Confined Monolayer Graphene Quantum Dots with Orbital and Valley Splittings. *Nano Lett.* **16**, 5798–5805 (2016).

20. J. Lee *et al.*, Imaging electrostatically confined Dirac fermions in graphene quantum dots. *Nat. Phys.* **12**, 1032–1036 (2016).

21. C. Gutiérrez, L. Brown, C.-J. Kim, J. Park, A. N. Pasupathy, Klein tunnelling and electron trapping in nanometre-scale graphene quantum dots. *Nat. Phys.* **12**, 1069–1075 (2016).

22. F. Ghahari *et al.*, An on/off Berry phase switch in circular graphene resonators. *Science*. **356**, 845–849 (2017).

23. A. Matulis, F. M. Peeters, Quasibound states of quantum dots in single and bilayer graphene. *Phys. Rev. B*. **77**, 115423 (2008).

24. J. F. Rodriguez-Nieva, L. S. Levitov, Berry phase jumps and giant nonreciprocity in Dirac quantum dots. *Phys. Rev. B*. **94**, 235406 (2016).

25. Additional supplementary text and data are available on Science Online.

26. M. Morgenstern, D. Haude, J. Klijn, R. Wiesendanger, Coulomb pseudogap caused by partial localization of a three-dimensional electron system in the extreme quantum limit. *Phys. Rev. B*. **66**, 121102 (2002).

27. S. Becker *et al.*, Probing Electron-Electron Interaction in Quantum Hall Systems with Scanning Tunneling Spectroscopy. *Phys. Rev. Lett.* **106**, 156805 (2011).

28. Y. J. Song *et al.*, High-resolution tunnelling spectroscopy of a graphene quartet. *Nature*. **467**, 185–189 (2010).

29. A. V. Filinov, M. Bonitz, Y. E. Lozovik, Wigner Crystallization in Mesoscopic 2D Electron Systems. *Phys. Rev. Lett.* **86**, 3851–3854 (2001).

30. C.-H. Zhang, Y. Joglekar, Wigner crystal and bubble phases in graphene in the quantum Hall regime. *Phys. Rev. B*. **75**, 1–6 (2007).

31. K. A. Guerrero-Becerra, M. Rontani, Wigner localization in a graphene quantum dot with a mass gap. *Phys. Rev. B*. **90**, 125446 (2014).

32. C. Dean *et al.*, Boron nitride substrates for high-quality graphene electronics. *Nat. Nano*. **5**, 722–726 (2010).

33. D. T. McClure *et al.*, Edge-State Velocity and Coherence in a Quantum Hall Fabry-P\'erot Interferometer. *Phys. Rev. Lett.* **103**, 206806 (2009).

34. J. Jang, B. M. Hunt, L. N. Pfeiffer, K. W. West, R. C. Ashoori, Sharp tunnelling resonance from the vibrations of an electronic Wigner crystal. *Nat. Phys.* **13**, 340–344 (2017).

35. M. F. Crommie, C. P. Lutz, D. M. Eigler, Imaging standing waves in a two-dimensional electron gas. *Nature*. **363**, 524–527 (1993).

36. M. F. Crommie, C. P. Lutz, D. M. Eigler, Confinement of Electrons to Quantum Corrals on a Metal Surface. *Science*. **262**, 218–220 (1993).

37. The uncertainty represents one standard deviation determined from a linear least squares fit to the data.

38. D. C. Elias *et al.*, Dirac cones reshaped by interaction effects in suspended graphene. *Nat Phys*. **7**, 701–704 (2011).

39. L. Susskind, Lattice fermions. *Phys. Rev. D*. **16**, 3031–3039 (1977).

40. R. Stacey, Eliminating lattice fermion doubling. *Phys. Rev. D*. **26**, 468–472 (1982).

**ACKNOWLEDGMENTS**

We thank Steve Blankenship and Alan Band for their contributions to this project, and we thank Mark Stiles for valuable discussions.

**Funding:** C.G., D.W., and F.G., acknowledges support under the Cooperative Research Agreement between the University of Maryland and the National Institute of Standards and Technology Center for Nanoscale Science and Technology, Grant No. 70NANB10H193, through the University of Maryland. C.L. and L.S.L. acknowledge support by the STC Center for Integrated Quantum Materials (CIQM) under NSF award 1231319. L.S.L. also acknowledges support by the U.S. Army Research Laboratory and the U.S. Army Research Office through the Institute for Soldier Nanotechnologies, under contract number W911NF-13-D-0001. K.W. and T.T. acknowledge support for the growth of hexagonal boron nitride crystals by the Elemental Strategy Initiative conducted by the MEXT, Japan and the CREST (JPMJCR15F3), JST.

**Author contributions:** C.G., D.W., and F.G performed the experiments. C.G. led the STM/STS data analysis. F.G. designed and fabricated the graphene device. J.A.S. designed and built the apparatus, and conceived of the study. C.L., J.F.R-N., and L.S.L performed the simulations. K.W. and T.T. grew the hBN crystals used in the graphene device. All authors contributed to writing the manuscript.
**Competing interests:** None.
**Data availability:** All data are available in the manuscript and supplementary materials.

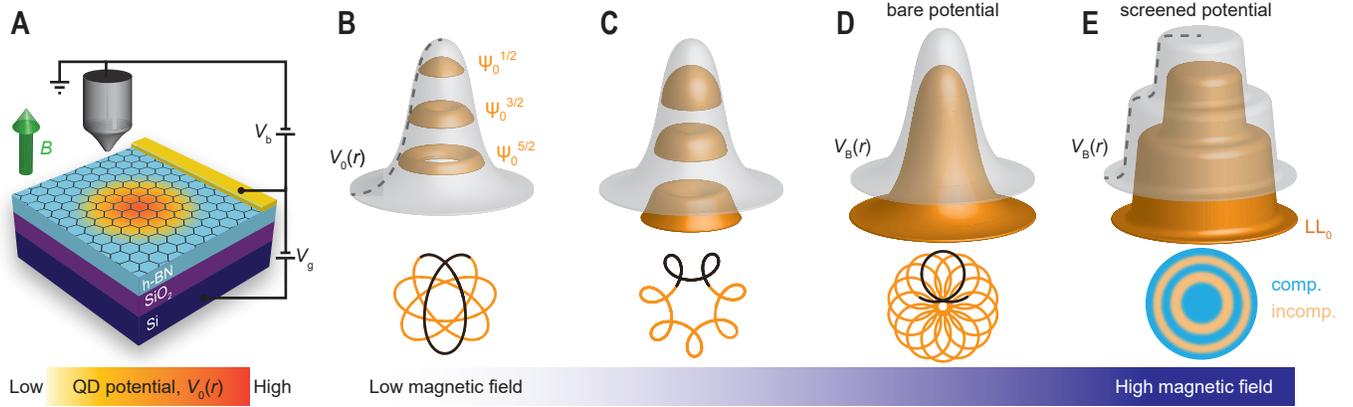

Figure 1

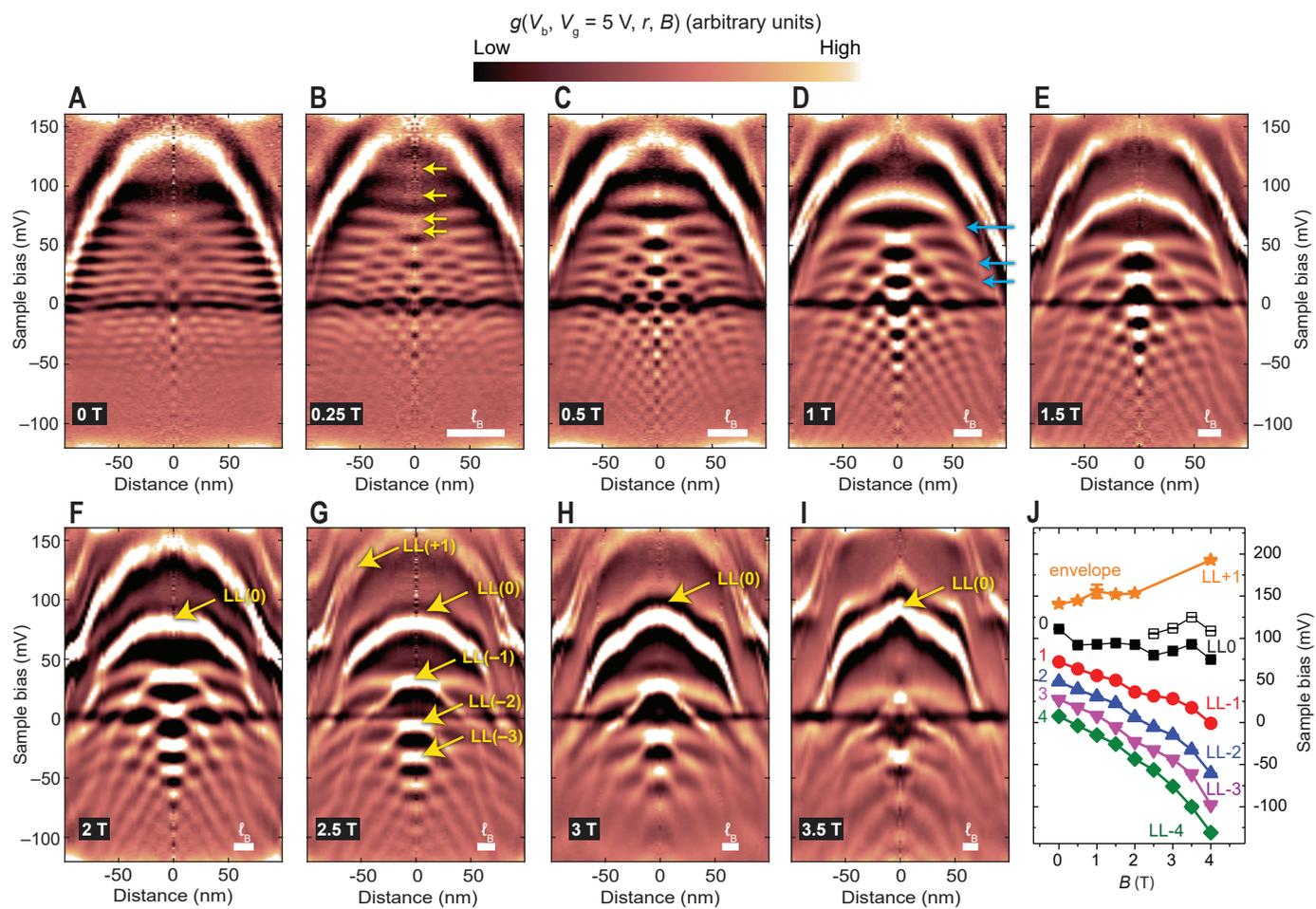

Figure 2

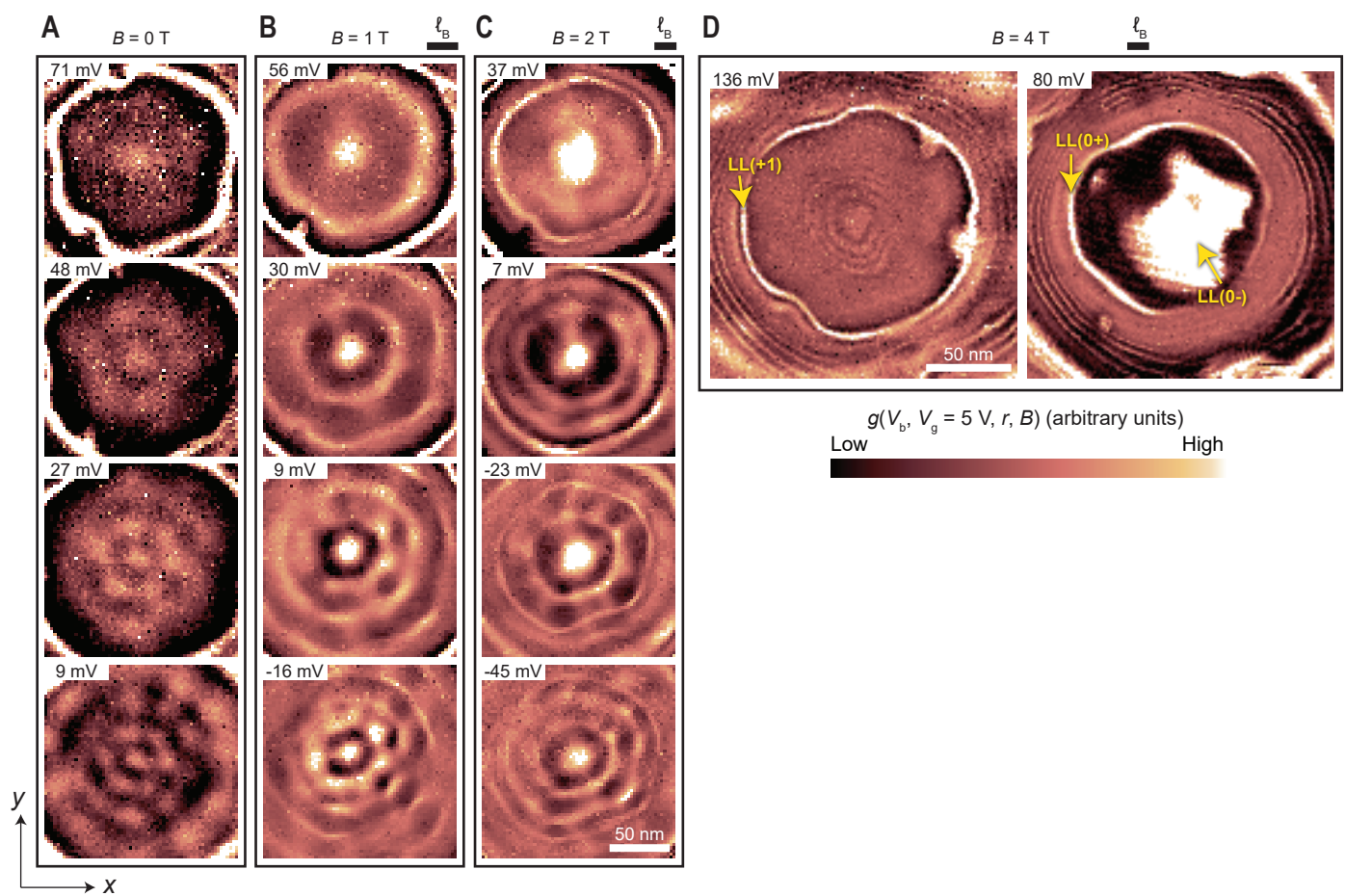

Figure 3

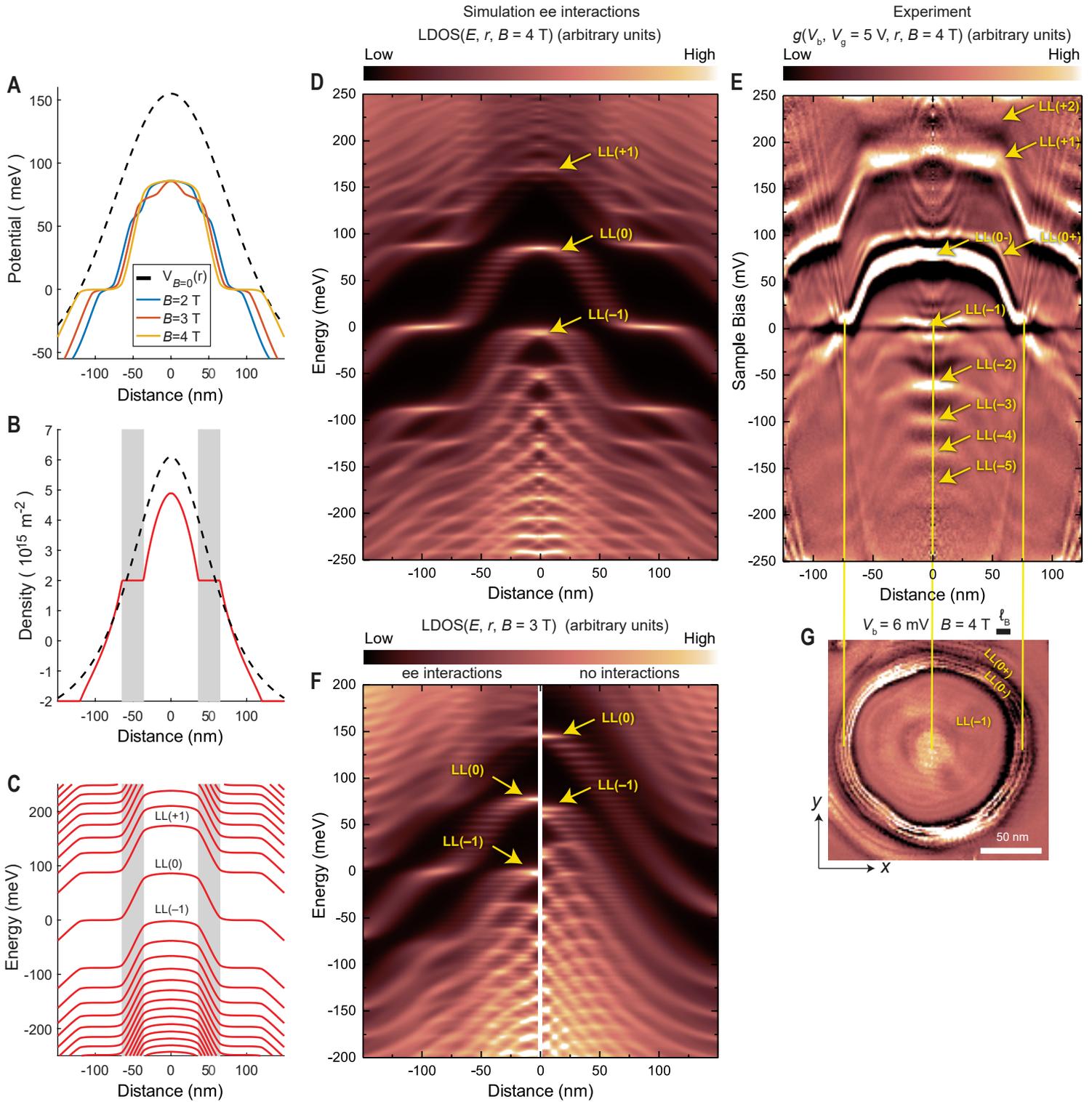

Figure 4

# Supplementary Materials for

## Interaction Driven Quantum-Hall Wedding cake-like Structures in Graphene Quantum Dots

Christopher Gutiérrez[1,2,*,†], Daniel Walkup[1,2,*], Fereshte Ghahari[1,2,*], Cyprian Lewandowski[3,*], Joaquin F. Rodriguez-Nieva[4], Kenji Watanabe[5], Takashi Taniguchi[5], Leonid S. Levitov[3], Nikolai B. Zhitenev[1], and Joseph A. Stroscio[1]

[1]Center for Nanoscale Science and Technology, National Institute of Standards and Technology, Gaithersburg, MD 20899, USA
[2]Maryland NanoCenter, University of Maryland, College Park, MD 20742, USA
[3]Department of Physics, Massachusetts Institute of Technology, Cambridge, MA 02139, USA
[4]Department of Physics, Harvard University, Cambridge, MA 02138, USA
[5]National Institute for Materials Science, Tsukuba, Ibaraki 305-0044, Japan

correspondence to:  joseph.stroscio@nist.gov

**This PDF file includes:**

    I.     Methods and sample fabrication

    II.    Enhanced Fermi velocity measurements

    III.   Modeling the wedding cake potential

    IV.   Estimating the width of the incompressible region

           Figures S1 and S2
           Captions for Movie S1

**Other Supplementary Materials for this manuscript includes the following:**

       Movie S1

---

[*] These authors contributed equally to this work.
[†] Present address: Quantum Matter Institute, University of British Columbia, Vancouver, British Columbia V6T 1Z4, Canada

### I. Methods and sample fabrication

Our graphene heterostructure device consists of monolayer graphene on 20 nm thick hexagonal boron nitride (hBN) on 285 nm $SiO_2$/Si. Details on this device and its assembly have been reported previously (*22*) using a transfer method described in Ref. (*32*). To summarize briefly, single crystals of hBN were exfoliated onto $SiO_2$/Si substrates where a suitably thick flake (20 nm) was selected for further processing. Separately, monolayer graphene flakes were exfoliated onto a stack consisting of polymethyl methacrylate (PMMA)/polyvinyl alcohol (PVA)/Si. PVA is water-soluble and acts as a sacrificial layer for delaminating the graphene and allowing it to be transferred onto the target hBN/$SiO_2$/Si using a micromanipulator. After transferring, Cr(1 nm)/Pd(10 nm)/Au(40 nm) electrical contacts, including two sets of radial guides for STM navigation, were deposited onto the sample using standard e-beam lithography processing. The final device is annealed for several hours in 5% $H_2$/95% Ar at 350 °C to remove any processing residues. The sample was annealed one final time in an ultra-high vacuum chamber at 350 °C for several hours prior to STM measurements.

The graphene quantum dots (QD) were made by ionizing impurities in the hBN substrate using the STM tip, as described in Ref. (*20*), creating a *p*-type QD embedded in an *n*-type background (see Fig. 1A of the main text). To achieve the specific nano-patterning in our device, the global backgate voltage is first set to $V_g = 30$ V and the STM tip is retracted by 2 nm. Next, the sample voltage bias (relative to the grounded STM tip) is ramped to 5 V and held for $t = 60$ s. The strong electric field just beneath the STM tip during the 5 V pulse ionizes impurities in the hBN which redistribute themselves to cancel out the field of the global backgate. Finally, the external gate is lowered to $V_g = 5$

V (corresponding to global *n*-doping for our device), whence the ionized impurities in the pulsed region act as a negative local embedded gate, resulting in a local *p*-doped region in the graphene. For all measurements in this report, the global backgate voltage was held fixed at $V_g = 5$ V (after the QD was created).

We probe the quantum states in the graphene QD by measuring the tunneling differential conductance, $g(V_b, V_g, r, B) = dI/dV_b$, as a function of tunneling bias, $V_b$, back gate potential, $V_g$, spatial position, $r$, and magnetic field, $B$. $dI/dV_b$ measurements were recorded via lock-in detection using an AC voltage of 2 mV at a frequency of 383 Hz with the STM feedback disengaged. All measurements were performed at $T = 4.3$ K.

Raw differential conductance curves measured on graphene quantum dots feature fine resonator state peaks superimposed on a large dispersive graphene background (*20*, *22*) (Fig. S1, black curve). This superposition makes it difficult to image all the salient features in the data on the same color scale, for example in 2D radial maps in the main text (Fig. 2). We thus follow refs. (*33*, *34*) and subtract a smoothly-varying background (red curve) from each $dI/dV_b$ curve and plot the residual (blue curve), as shown in Fig. S1. The smoothly-varying background for each $dI/dV_b$ curve is calculated by Gaussian smoothing the original data with a FWHM = 28.3 meV.

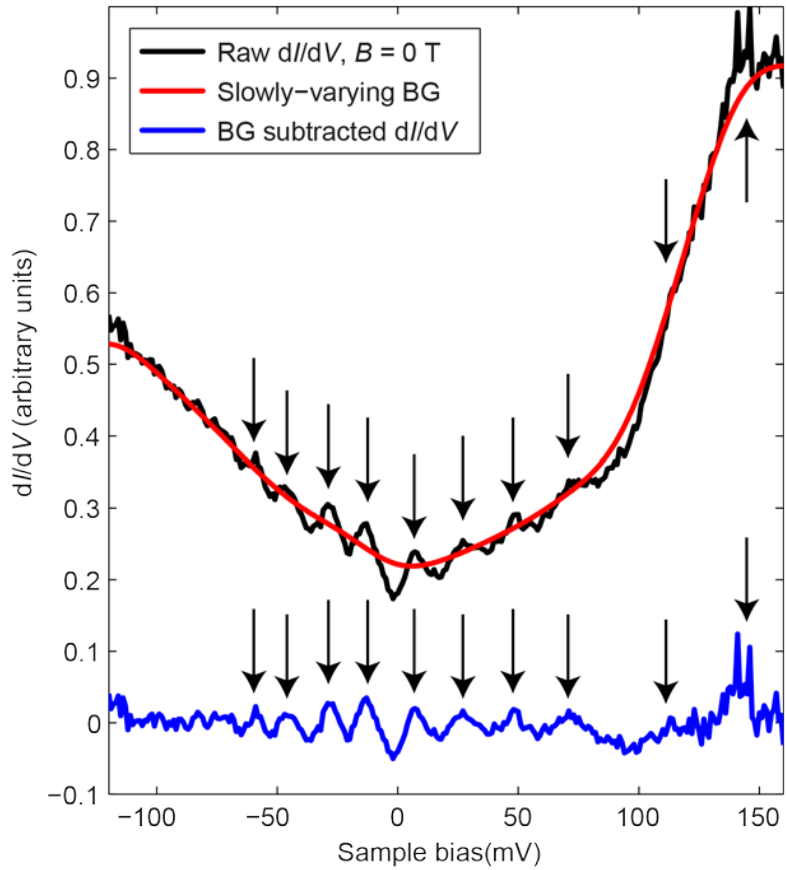

**Figure S1: Original and background-subtracted differential conductance**. Original $dI/dV_b$ spectra (black line) display a slowly-varying and dispersive graphene background (BG) (red line). In order to enhance the salient features in the LDOS, such as the QD states indicated by the arrows, the BG is subtracted from the original $dI/dV_b$ producing the blue curve, which preserves the positions of the QD resonance peaks (black arrows).

## II. Enhanced Fermi velocity measurements

Close inspection of the LLs in Fig. 2J shows that the LL energy spacings are larger than expected from theory (Eq. S1) and increase with increasing magnetic field. From Eq. S1 we attribute this increase to an effective enhancement of the Fermi velocity, $v_F$, which we extract from our experimental spectroscopic differential conductance maps using two methods: (1) At low fields ($B < 2.5$ T), we employ Fourier transforms of the radial $dI/dV$ maps, $\tilde{g}(q_x, q_y, V_b, V_g, B)$, to analyze quasiparticle interference (QPI) patterns in order to extract the graphene dispersion; and (2) at higher fields ($B > 1.5$ T), we use the peak positions of the characteristic graphene Landau level energies. The result of the analysis shows the Fermi velocity increasing with increased applied magnetic field.

### 1. QPI analysis

Surface defects and potential boundaries act as scattering sites for 2D electron gases (2DEGs) (*35*, *36*), whereby the scattered electrons interfere with each other and appear as standing waves in the local density of states (LDOS) with a characteristic scattering wavevector, **q** = **k**$_f$ - **k**$_i$, that connects two points on a constant energy surface. The Dirac fermions within our graphene QD form quasi-bound states due to Klein scattering at the walls of the potential boundary and appear as circular standing waves (*18–22*). At low energies, the graphene bandstructure is composed of linear Dirac cones at the K and K' points in the first Brillouin zone (Fig. S2A). The constant energy contours (CECs) are circles of radius $k(E)$, where the momentum $k$ depends on energy $E$ according to the graphene dispersion $E(k) = \hbar v_F k + E_D$, where $E_D$ is the Dirac point (Fig. S2A). The maximum scattering wavevector, q, of the circular QPI patterns is then given by the diameter of the CEC, $q = 2k$ (Fig. S2A, right). Figure S2B displays a tomographic slice of the fast Fourier transform (FFT) of the experimental $dI/dV$ map, $\tilde{g}$, recorded at $B = 0.5$ T. A linear dispersion is clearly observed, with a slope (red line, linear fit) given by $\hbar v_F/2$. Note that the Fermi energy ($V_b = 0$) cuts through the graphene valence band, confirming that the graphene QD is $p$-doped at its center.

### 2. Landau level analysis

In graphene (and other Dirac materials), the Landau level energies, $\varepsilon_N$, are unevenly spaced and given by the expression

$$\varepsilon_N = v_F\sqrt{2e\hbar B}\left[\text{sgn}(N)\sqrt{|N|}\right], N \in \mathbb{Z} \qquad (S1)$$

where $N$ is the (integer) Landau level index, $v_F$ is the graphene Fermi velocity, $e$ is the elementary charge, $\hbar$ is Planck's constant divided by $2\pi$, $B$ is the magnetic field, and sgn($N$) is the sign of the Landau level index. Figure S1C displays a $dI/dV$ spectra recorded at the center of the GQD ($r = 0$ nm) at $B = 3$ T and displays strong peaks that correspond to the large density of states at the highly-degenerate Landau levels. The Fermi velocity is then calculated from linear fits of the LL energy, $\varepsilon_N$, versus the bracketed term involving the LL index in Eq. S1 (Fig. S2C, inset).

### 3. Discussion

Combining the measured Fermi velocities using the two methods, we find that there is an enhancement of $v_F$ with increasing magnetic field (Fig. S2D). (The two data points at $B = 4$ T correspond to LL spectra measured inside and outside the graphene QD.) A linear fit of the combined data yields $v_F = v_0 + \alpha B$, where $v_0 = (1.017 \pm 0.022) \times 10^6$ m/s and $\alpha = 0.124 \pm 0.011$ ms$^{-1}$T$^{-1}$ (*37*). We note that Fermi velocity renormalization in graphene has been observed previously at low carrier densities ($n < 10^{12}$ cm$^{-2}$) near the Dirac point and was attributed to electronic interactions (*14*, *38*). This is in agreement with our observations of decreasing density, for example, at $B = 4$ T (Fig. 4E main text), the density is $n = \frac{E_D^2}{\pi \hbar^2 v_F^2} \approx 0.23 \times 10^{12}$ cm$^{-2}$ (for $E_D \approx 92$ meV, $v_F \approx 1.64 \times 10^6$ m/s). This is also consistent with increased electron-electron (*ee*) interactions with applied magnetic fields, as evidenced by the increase in the observed Fermi velocity, the kinks in the 'wedding cake' spatial structure of the LLs (Fig. 2 and Fig. 4E main text), and the need for self-consistent potentials to accurately simulate the data in applied magnetic fields.

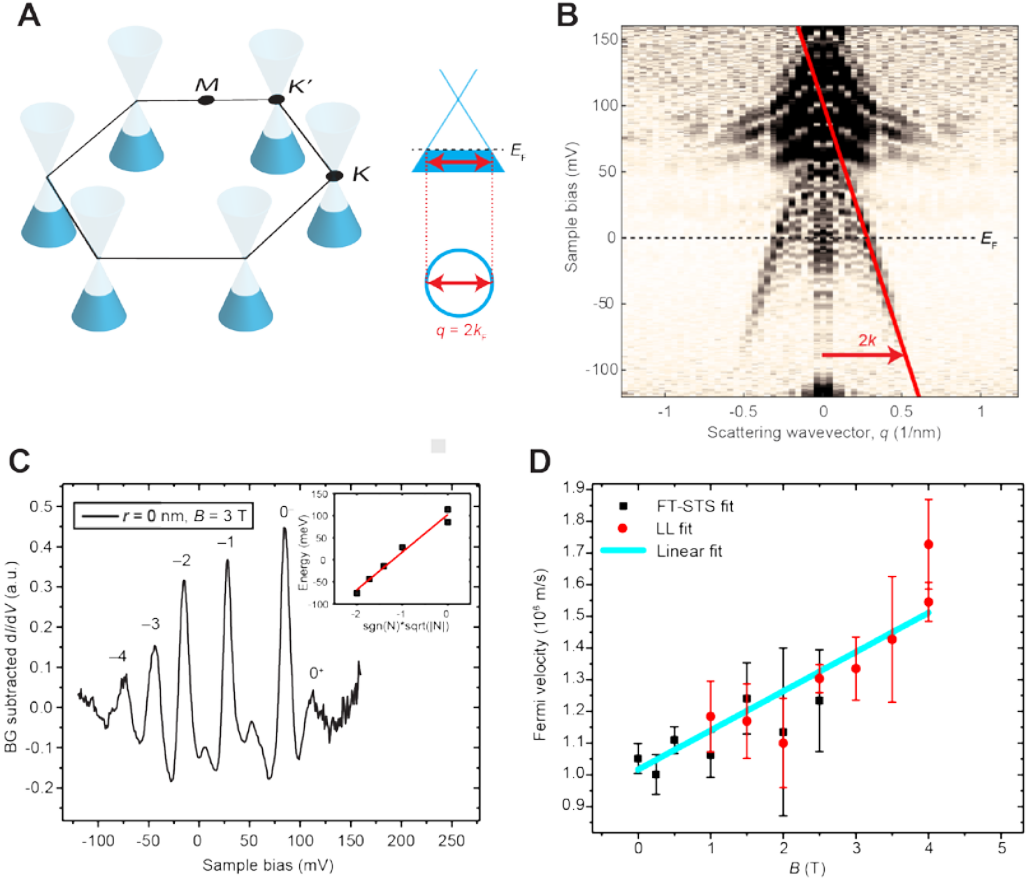

**Figure S2: Magnetic field-dependence of the Fermi velocity.** (**A**) Schematic of the low-energy band structure of graphene, displaying Dirac cones at the ***K/K'*** points. Right panel: Side-view of a single Dirac cone with the Fermi energy, $E_F$, crossing the valence band, signifying *p*-doping. The maximum scattering wavevector (red arrow) has a magnitude $q = 2k$. (**B**) Fourier transform scanning tunneling spectroscopy (FT-STS) map recorded at $B = 0.5$ T. A linear fit of the conical quasi-particle interference pattern (red line) yields a slope directly proportional to the graphene Fermi velocity. (**C**) *dI/dV* spectra recorded at $B = 3$ T, displaying sharp resonances at the graphene LLs. A slowly-varying, dispersive graphene background has been removed to highlight the salient features. Inset: A plot of the LL energies versus the LL index *N* and linear fit (red line) proportional to $v_F$. (**D**) A combined plot of the measured Fermi velocities as a function of magnetic field using the two methods displayed in (B) and (C). The Fermi velocity is seen to increase with increasing field, coinciding with the decrease in density and increasing importance of electron-electron (*ee*) interactions in the quantum dot system. Error bars represent one standard deviation uncertainties from linear least-squares fits to the QPI and LL plots as shown in (B) and (C), respectively.

### III. Modelling the Wedding Cake potential

Here we describe the approach used to model the experimental data. Our analysis proceeds in two steps. First, an effective electrostatic potential which accounts for screening and Coulomb repulsion is calculated. Second, this potential is used as an input for the one-particle Dirac equation to produce LDOS maps shown in Fig. 4.

1. <u>Determining the self-consistent potential and charge density</u>

We consider graphene's Dirac electrons in the quantum Hall regime and in the presence of an external electrostatic potential $V_{\text{ext}}(r)$. We assume that the $V_{\text{ext}}(r)$ spatial variation is slow on the scale of the magnetic length $l_B$. The total energy of the system is a sum of contributions from the kinetic energy due to the cyclotron motion, the potential energy due to $V_{\text{ext}}(r)$, and the Coulomb energy due to electron-electron (*ee*) repulsion. As discussed in the main text, the wedding cake-like structure results from the competition between the kinetic energy (i.e. filling the lowest possible LL) and the potential energy due to *ee* repulsion and the external potential. These competing behaviors are captured by the energy functional introduced and discussed in the main text (also, see (*4*)):

$$E[n] = \int d^2 r \left( K[n(r)] + V_{\text{ext}}(r)n(r) + \frac{1}{2}\int d^2 r' V_{\text{ee}}(|\mathbf{r}-\mathbf{r}'|)n(r)n(r') \right). \quad (S2)$$

Here $n(r)$ is the graphene charge density at position $\mathbf{r}$, and $K[n(r)]$ is the kinetic energy due to the Landau levels:

$$\frac{\delta K}{\delta n(r)} = \varepsilon_N, \quad \left(N - \frac{1}{2}\right)n_{LL} < n(r) < \left(N + \frac{1}{2}\right)n_{LL}. \quad (S3)$$

Here $n_{LL}$ is the density of a filled LL $n_{LL} = g/2\pi l_B^2$, where $g = 4$ is the LL spin-valley degeneracy. The energy of the $N$th LL is:

$$\varepsilon_N = \hbar\omega_c \text{sgn}(N)\sqrt{|N|}, \quad \omega_c = v_F\sqrt{\frac{2eB}{\hbar}}. \tag{S4}$$

The $ee$ interaction is given by $V_{ee}(r) = \frac{\tilde{e}^2}{r}$. Here $\tilde{e}$ is the screened electron charge, $\tilde{e}^2 = \frac{e^2}{4\pi\epsilon_0 \kappa_{ave}}$ with $\kappa_{ave}$ an average dielectric constant and $\epsilon_0$ the vacuum permittivity.

The quantum dot is defined by the potential induced by localized charges in the ionized region of the substrate $V_{ext}(r)$, offset by an electrostatic potential induced by uniform charge distribution $-n_g$ at the back gate, both of which contribute to the external potential $V_{ext}(r)$ in Eq. S2. Naturally, the total charge at the gates is much greater than the total charge in the ionized region, and the graphene-gate distance is much smaller than the size of the graphene flake. As a result, we assume that graphene screens the charge at the gate. We therefore set $n(r) = n_g + \delta n(r)$, where $\delta n$ is the charge induced by $V_{ext}(r)$. The functional can then be expressed in terms of $\delta n(r)$ as:

$$E[\delta n] = \int d^2r \left( K[n_g + \delta n(r)] + \tilde{V}_{ext}(r)\delta n(r) + \frac{1}{2}\int d^2r' V_{ee}(|\boldsymbol{r} - \boldsymbol{r}'|)\delta n(r)\delta n(r') \right), \tag{S5}$$

where we subtracted a constant that does not depend on $\delta n$.

For the localized charges in the ionized region, we use the potential corresponding to a point-like charge potential:

$$\tilde{V}_{\text{ext}}(r) = \frac{\tilde{V}_{\text{ext},0}}{\sqrt{1 + \frac{r^2}{r_{\text{ext}}^2}}} \tag{S6}$$

with parameters $\tilde{V}_{\text{ext},0} = 1450$ meV and $r_{\text{ext}} = 85$ nm to reproduce the observed charge density and its spatial extent.

To minimize Eq. S5, we first note that the term $K[\delta n(r)]$ introduces non-linearity to the functional $E[\delta n]$. Nevertheless, $E[\delta n]$ is a concave function of $\delta n(r)$ and, as such, is amenable to gradient descent. We begin by using a trial solution $\delta n_0(r) = 0$. For each $r$, we compute the direction of ascending $E$:

$$\frac{\delta E}{\delta n(r)} = \frac{\delta K}{\delta n}[n_g + \delta n_k(r)] + \tilde{V}_{\text{ext}}(r) + \int d^2r' V_{ee}(|\mathbf{r} - \mathbf{r}'|)\delta n_k(r'). \tag{S7}$$

The density profile is then updated using

$$\delta n_{k+1}(r) = \delta n_k(r) - \frac{\delta E}{\delta n(r)}[\delta n_k]h, \tag{S8}$$

where $h$ is a small step size in the direction of descending energy. The procedure is continued until the minimum of the functional in Eq.(S5) is reached:

$$\frac{\delta K}{\delta n}[n_g + \delta n(r)] + \tilde{V}_{\text{ext}}(r) + \int d^2r' V_{ee}(|\mathbf{r} - \mathbf{r}'|)\delta n(r') = 0. \tag{S9}$$

From the solution of Eq. S7, we define the effective potential $V_B(r)$:

$$V_B(r) = V_{\text{ext}}(r) + \int d^2r' V_{ee}(|\mathbf{r} - \mathbf{r}'|)\delta n(r'), \tag{S10}$$

which is used as the input of the one-particle Dirac equation (see next section).

The charge density $n(r)$ is shown in Fig. 4B and the effective potential $V_B(r)$ for several magnetic field values is presented in Fig. 4A of the main text. In Fig. 4C we plot the effective potential $V_B(r)$ displaced by consecutive LLs energies, Eq.(S4), for $N = -14, -13, \ldots, 10$. The compressible and incompressible regions corresponding to the plateaus and concave elements of the screened potential are described in the main text.

## 2. Solving the Dirac equation

We consider the Dirac equation for radially confined electrons in the presence of a uniform magnetic field:

$$[v\,\boldsymbol{\sigma}\cdot\boldsymbol{q} + V_B(r)]\,\Psi(\boldsymbol{r}) = \varepsilon\,\Psi(\boldsymbol{r}) \tag{S11}$$

Here $\boldsymbol{q}$ is the kinematic momentum with components $q_{x,y} = -i\hbar\partial_{x,y} - eA_{x,y}$ and $q_z = 0$. The approach presented here follows the same reasoning as given in Ref. (*24*), but we reproduce and expand the discussion here for completeness.

Due to the rotational symmetry of the potential $V_B(r)$ we use the axial gauge $A_x = -By/2, A_y = Bx/2$. The eigenstates of Eq. S11 can be then expressed using the polar decomposition ansatz,

$$\Psi_m(r,\theta) = \frac{e^{im\theta}}{\sqrt{r}}\begin{pmatrix} u_1(r)e^{-\frac{i\theta}{2}} \\ iu_2(r)e^{\frac{i\theta}{2}} \end{pmatrix} \tag{S12}$$

with $m$ a half-integer number. This decomposition allows to rewrite Eq. (S11) as

$$\begin{pmatrix} V_B(r) - \varepsilon & \partial_r + m/r - Br/2 \\ -\partial_r + m/r - Br/2 & V_B(r) - \varepsilon \end{pmatrix}\begin{pmatrix} u_1 \\ u_2 \end{pmatrix} = 0 \tag{S13}$$

Connection with an experimental measurement of conductance $g = dI/dV$ is provided via a local density of states (LDOS) $g \propto D(\epsilon, r)$. The quantity $D(\epsilon, r)$ can be conveniently written as the sum of $m$-state contributions $D(\epsilon, r) = \sum_m D_m(\epsilon, r)$, with

$$D_m(\epsilon, r) = \sum_\alpha \frac{|u_\alpha(r)|^2}{r} \delta(\epsilon - \epsilon_\alpha). \tag{S14}$$

Here $\alpha$ labels the radial eigenstates of Eq. S13 for fixed $m$.

When discretizing a Dirac equation on a lattice one encounters the problem of Fermion doubling. One standard approach is to use a forward-backward difference scheme (*39*, *40*) for approximating the partial derivatives in Eq. (S13)

$$\partial_r u_1 \approx \frac{u_1(r) - u_1(r-h)}{h}, \qquad \partial_r u_2 \approx \frac{u_2(r+h) - u_2(r)}{h} \tag{S15}$$

where $h = \frac{L}{N-1}$ corresponds to the discretization step size for a system of size $L$ and $N$ lattice sites. This requires us to specify boundary conditions on $u_1(0)$ and $u_2(L)$, which, to preserve the hermiticity of the Hamiltonian, are taken as $u_1(0) = 0$ and $u_2(L) = 0$. The latter boundary condition does not carry any consequence in context of the LDOS maps. On the other hand, the vanishing of $u_1(0)$ does matter as it forces the LDOS to vanish at the origin – an unphysical condition. As a remark, we note that this does not impact the local density of states a few step sizes away from the origin.

In order to produce spectral maps free of this artifact, while preserving hermiticity of the discretized Hamiltonian and avoiding the fermion doubling problem, we employ a simple trick: we compute the local density of states using both forward-backward and backward-forward difference schemes and combine the two results. In the backward-forward scheme the partial derivatives from Eq. S13 take the form:

$$\partial_r u_1 \approx \frac{u_1(r+h) - u_1(r)}{h}, \qquad \partial_r u_2 \approx \frac{u_2(r) - u_2(r-h)}{h} \tag{S16}$$

The simulation was run on a lattice consisting of $N = 600$ sites and a system size $L = 430$ nm. Level broadening was chosen as $\gamma = 2.4$ meV. The range of angular momenta summed was estimated as to include all states that give rise to physical features in the LDOS of region of interest. As in earlier works, the contribution of spurious states present due to a finite system size were excluded.

### IV. Estimating the width of the incompressible region

The width of the observed incompressible ring can be estimated from the functional in Eq. S5 following the approach of Ref. (*5*). In the absence of the kinetic energy term, $K[n(r)]$, the system is fully compressible and Eq. S5 predicts a smooth charge density profile that spans the entire QD (the dashed line in Fig. 4B). Once $K[n(r)]$ is restored, Eq. S5 predicts flat regions as illustrated in Fig. 4B, with the charge density profile shown as the solid line. These regions correspond to incompressible rings in $n(r)$, formed between LLs crossings with the Fermi level. The incompressible region size can be estimated by considering a dipolar strip of width $l$ and optimizing $l$ to minimize Coulomb repulsion between graphene electrons. Qualitatively the resulting value of $l$ is such that the built-in electric field within the strip, $\tilde{e}Ea \sim \tilde{e}^2 \frac{dn}{dr} l^2$ matches the LL separation, $\Delta\varepsilon_{\text{LL}}$. Here we provide a derivation of the Eq. 4 used in the main text, which provides a quantitative estimate of the dipolar strip width.

Following Ref. (5), we write an electric potential of a two-dimensional electron system (2DES) containing the incompressible region in terms of suitably chosen harmonic functions:

$$\phi(z) = Im\left[\frac{u_1}{\pi}\ln\left((z^2-a^2)^{\frac{1}{2}}+z\right) + u_2(z^2-a^2)^{1/2}z + u_3 z\right], \quad y \geq 0 \quad (S17)$$

where $z$ is a complex variable $z = x + iy$ with $x$ replacing $r$ and $y$ denoting the vertical coordinate perpendicular to the 2D layer. Here, following (5), we consider a quasi-1D linear geometry in which potential and density depend on one of the Cartesian coordinates in the plane but do not depend on the other coordinate. This corresponds to the limit of the incompressible ring in our QHE droplet being much narrower than the droplet radius. We assume that the incompressible region occurs at $-a < x < a$.

We note that, while the main ingredients in our problem are the same as in that analyzed in Ref. (5), there is a slight difference in the geometry that leads to extra numerical factors in the final result. Namely, Ref. (5) considers a 2DES with proximal top gates parallel to it and a dielectric beneath it, and obtains an incompressible strip of width greater than the distance to the gates. Here, in contrast, the incompressible strip width is small compared to the distance to the back-gate, and therefore we have to analyze a 2DES with dielectric beneath it and vacuum above it. This problem is equivalent to a more symmetric problem with a dielectric on both sides of the 2DES of an effective dielectric constant $\kappa_{ave} = (\kappa + 1)/2$, where $\kappa$ is the permittivity beneath the 2DES. In this case, the potential of a charge plane $\sigma(x)$ representing 2DES has mirror symmetry with respect to the plane, given by Eq. S17 at $y > 0$ and by an identical function beneath the plane such that the potential is overall $y/-y$ symmetric. For clarity in what follows we perform the calculation in terms of a screened electron charge $\tilde{e}^2 = \frac{e^2}{4\pi\epsilon_0 \kappa_{ave}}$.

The charge density $\sigma(x)$ is related to $\phi(x,y)$ by the Gauss' law:

$$\sigma(x) = -\frac{1}{2\pi}\frac{\partial \phi}{\partial y}\bigg|_{y=0^+}. \tag{S18}$$

In the incompressible region the contribution to charge density due to the first two terms of Eq. (S17) vanishes. This fixes the relation $u_3$ in terms of the LL density as:

$$u_3 = 2\pi \tilde{e} n_{LL}. \tag{S19}$$

Next, the condition that the tangential electric field must vanish at the boundary of the incompressible region yields a relation between coefficients $u_1$ and $u_2$:

$$u_1 = -\pi u_2 a^2. \tag{S20}$$

Combining Eq. S20 with Eq. S18 gives the charge density:

$$\sigma(x) = \tilde{e} n_{LL} + \frac{u_2}{\pi} \begin{cases} \sqrt{x^2 - a^2} & \text{for } x > a \\ 0 & \text{for } -a < x < a \\ -\sqrt{x^2 - a^2} & \text{for } -a < x \end{cases}. \tag{S21}$$

Finally, since the dependence at $x \gg a$ is linear with $x$, we can express $u_2$ in terms of the charge density gradient $\frac{d\sigma}{dx}$ in the compressible region outside the incompressible strip:

$$\frac{d\sigma}{dx}\bigg|_{x \gg a} = \frac{u_2}{\pi}. \tag{S22}$$

This expression allows us to relate $u_2$ to the $\frac{dn}{dx}$ value, which was calculated numerically for a compressible droplet (see dashed line in Fig. 4B):

$$\tilde{e}\frac{dn}{dx}\bigg|_{x=0} = \frac{u_2}{\pi}. \tag{S23}$$

Using the fact that the drop of the electrostatic potential across the dipolar incompressible strip is $\Delta\varepsilon_{LL}/\tilde{e}$, we get:

$$\tilde{e}u_1 = -\Delta\varepsilon_{LL}, \tag{S24}$$

where $\Delta\varepsilon_{LL} = \hbar v_F \sqrt{2}/l_B$ is the average LL energy spacing.

Combining Eqs. S21, S23 and S24, we obtain the width of the incompressible region

$$l = \left(\frac{4\Delta\epsilon_{LL}}{\pi^2 \tilde{e}^2 \frac{dn}{dx}\big|_{x=0}}\right)^{\frac{1}{2}} \approx 34\ nm, \tag{S25}$$

as quoted in Eq. 4 in the main text and $\tilde{e}^2 = \frac{e^2}{4\pi\epsilon_0 \kappa_{ave}}$. Here we used the average dielectric constant of the substrate and vacuum $\kappa_{ave} = (\kappa_{SiO} + 1)/2$, the Landau level spacing $\Delta\varepsilon_{LL} = \hbar v_F\sqrt{2}/l_B$ for the observed levels $N$=0 and -1, and the density gradient $\frac{dn}{dr} \approx 3.9 \times 10^{22} m^{-3}$ for the compressible droplet in the middle of the would-be incompressible region (dashed line in Fig. 4B). We note that Eq. 20 in Ref. (*5*) has a similar form, except for a numerical prefactor 2 instead of the prefactor 4. The two expressions are in agreement in the limit of the substrate's dielectric constant being much greater than that of vacuum as assumed in Ref. (*5*). In the case that the substrate's dielectric constant is comparable to that of the vacuum, then Eq. S25 is more accurate.

The estimate in Eq. S25 is slightly greater than the width inferred from our measurement results shown in Fig.4E of the main text. The small discrepancy can be attributed to the result of Ref. (5), derived for LL spacing $\Delta\varepsilon_{LL}$ much smaller than the external potential, being used in the regime when $\Delta\varepsilon_{LL}$ is not small on the $V_B(r)$ scale. In this case, in contrast to Ref. (5), the incompressible strip width $l$ is not small compared to the QHE droplet radius. These estimates may also be affected by renormalization of $\kappa_{ave}$ due to interband polarization, which will be discussed elsewhere.

**Movie S1.** Experimental differential conductance, $g(V_b, V_g, r, B)$, maps of the local density of states of the graphene QD in the *x-y* plane as a function of sample bias, $V_b$, indicated in the top center of the movie. The movie has four quadrants showing four different magnetic fields, 0, 1, 2, and 3 T. Various QD states are seen coming in at different bias, corresponding to the states observed in Fig. 2 of the main text. A smooth background was subtracted from each *dI/dV* spectra to remove the graphene dispersion in order to visualize all features on a single color scale.